\documentclass[twocolumn,showpacs,pra,aps]{revtex4-1}
\usepackage{graphicx}
\usepackage{epstopdf}
\usepackage{amsmath}
\usepackage{amssymb}
\usepackage{epsfig}
\usepackage{amsthm}
\usepackage{color}

\usepackage{bbm,mathrsfs}
\usepackage{graphicx}
\usepackage{amsfonts,amsmath}
\usepackage{amsthm}
\usepackage{times}

\def\textbf#1{{\bf #1}}
\def\be{\begin{equation}}
\def\ee{\end{equation}}
\def\ben{\begin{eqnarray}}
\def\een{\end{eqnarray}}
\def\eea{\end{array}}
\def\bea{\begin{array}}
\newcommand{\Tr}[0]{\mathrm{Tr}}

\newcommand{\bei}{\begin{itemize}}
\newcommand{\eei}{\end{itemize}}
\newcommand{\ket}[1]{|#1\rangle}

\def\C{{\mathbb C}}
\def\one{\leavevmode\hbox{\small1\normalsize\kern-.33em1}}

\newtheorem{lemma}{Lemma}

\usepackage{changes}

\definecolor{myurlcolor}{rgb}{0,0,0.7}
\definecolor{myrefcolor}{rgb}{0.8,0,0}
\usepackage{hyperref}
\hypersetup{colorlinks, linkcolor=myrefcolor,
citecolor=myurlcolor, urlcolor=myurlcolor}

\begin{document}
\title{Bounding the persistency of the nonlocality of W states}

\author{P\'eter Divi\'anszky}
\author{R\'eka Trencs\'enyi}
\author{Erika Bene}
\author{Tam\'as V\'ertesi}
\affiliation{Institute for Nuclear Research, Hungarian Academy of Sciences, H-4001 Debrecen, P.O. Box 51, Hungary}

\begin{abstract}

The nonlocal properties of the W states are investigated under
particle loss. By removing all but two particles from an $N$-qubit
W state, the resulting two-qubit state is still entangled. Hence,
the W state has high persistency of entanglement. We ask an
analogous question regarding the persistency of nonlocality
introduced in [Phys. Rev. A 86, 042113]. Namely, we inquire what
is the minimal number of particles that must be removed from the W
state so that the resulting state becomes local. We bound this
value in function of $N$ qubits by considering Bell nonlocality
tests with two alternative settings per site. In particular, we
find that this value is between $2N/5$ and $N/2$ for large $N$. We
also develop a framework to establish bounds for more than two
settings per site.
\end{abstract}

 \pacs{}

\maketitle

\section{Introduction}
\label{intro}

Entanglement and nonlocality are two manifestations of quantum
correlations, both are essential ingredients of quantum theory.
They also play an important role in the field of quantum information
\cite{nielsenchuang}. For instance, entanglement is in the heart
of quantum teleportation and plays a crucial role in quantum
algorithms \cite{horodecki,geza}. Nonlocality, on the other hand, witnessed by the violation of Bell inequalities~\cite{bell}, reduces communication complexity \cite{buhrmanreview,buhrman} and enables device-independent quantum information protocols~\cite{bellreview,valerio}. These protocols do not rely on a detailed knowledge of the internal working of the experimental devices used, thereby allowing secure cryptography involving untrusted devices~\cite{qkd} or expansion of secure random numbers~\cite{random1,random2}.

The quantification of entanglement and nonlocality in the
bipartite case (i.e., the case when two parties share an entangled
state) is relatively well understood. However, the multipartite
case is much less explored. This is mainly due to the rapidly
growing complexity of the problem with the number of parties. For
instance, for bipartite pure states, there is a unique measure of
entanglement, however, for three or more parties, this is not true
any more (see e.g., Refs.~\cite{plenio,szalay}). Concerning
nonlocality, there is a single tight Bell inequality for two
binary settings per party, the Clauser-Horne-Shimony-Holt (CHSH)
one \cite{bell,chsh}. However, moving to three parties, the number
of tight Bell inequalities already becomes 42 \cite{sliwa}. In
fact, determining all Bell inequalities for growing number parties
is an NP-hard problem \cite{pitowski,NP}. Bipartite and tripartite
quantum nonlocality are different from a fundamental point of view
as well. While Gleason's theorem can be extended to the bipartite
scenario, however, there remains a gap between the quantum and
Gleason's correlations in the tripartite scenario \cite{acinObs}.

One of the most famous multipartite quantum states is the W state
\cite{Wstate} playing a crucial role in the physics of the
interaction between light and matter. Up to now, W states have been
prepared in lots of experiments, e.g., photonic experiments can
generate and characterize six qubit entangled states~\cite{Wexp1}.
More recently, genuine 28-particle entanglement was detected in a
Dicke-like state (a generalization of the W state) in a
Bose-Einstein condensate~\cite{Wexp2}.

W states are very robust to losses, hence suited to quantum
information applications such as quantum memories \cite{qmemory}.
For instance, by tracing out all but two parties from an $N$-party
W state, the remaining two-qubit state is still entangled, no
matter how big $N$ is. Hence, the W state shows a high persistency
of entanglement against particle loss (actually, the highest
possible persistency among $N$ qubit states~\cite{buzek}).

On the other hand, one may wonder how robust the nonlocality of
the W state is with respect to particle loss. In order to quantify
this, Ref.~\cite{persistency} introduces the persistency of
nonlocality of $N$-party quantum states $\rho$, $P_{NL}(\rho)$,
which is the minimal number of particles to be removed for
nonlocal quantum correlations to vanish. Ref.~\cite{persistency}
investigates this measure for various classes of multipartite
states including W states up to $N=7$ with two settings per site.
In this paper, we bound this value both from above and from below
for any $N$-qubit W states. To do this, we refine the definition
of $P_{NL}(\rho)$ to account for Bell nonlocality involving $m$
settings per party. This quantity will be called $P_{NL}^m(\rho)$.
Clearly, in the limit of large $m$, $P_{NL}^m(\rho)$ tends to
$P_{NL}(\rho)$. Our main result concerns the case of $m=2$ and we
prove the bounds of $2N/5\le P_{NL}^2(W_N)\le N/2$ for $N$ large,
featuring a relatively small gap between the upper and lower
bounds. The lower bound is based on an explicit construction of a
class of Bell inequalities. We also give a numerical framework to
put reliable lower bounds on $P_{NL}^m(W_N)$ beyond two settings
(up to $m=6$) and a tractable number of $N$ parties. This
numerical study supports that our analytical lower bounds for
$P_{NL}^2(W_N)$ are tight. There are recent papers which discuss
the robustness of Dicke states \cite{dicke} (and in particular the
W state) to various types of noises \cite{rafael,tomer,sohbdi}.
Lower bounds also follow from these papers for the value of
$P_{NL}^2(W_N)$. In particular, we obtain considerable improvement
over the lower bounds presented in Ref.~\cite{tomer}.

Notably, the persistency of nonlocality also gives a
device-independent bound on the persistency of entanglement
introduced in Ref.~\cite{briegel}. Other device-independent
approaches to quantify multipartite entanglement including W
states appeared in Refs.~\cite{DIW,tomoW,numW}. On the other hand,
multipartite W states are promising candidates to close the
detection loophole~\cite{eberhard} in multipartite Bell
tests~\cite{Wdetloophole}. Such Bell violations would complement
the experimental loophole-free violations obtained recently in the
bipartite case~\cite{loopholefree}.

The structure of the paper is as follows. Section~\ref{setup}
introduces notation. Section~\ref{UB} proves a simple upper bound
on the persistency of nonlocality $P^m_{NL}$ for W states and in general for any permutationally symmetric state with two
settings per party (case $m=2$). On the other hand, section~\ref{LB} presents lower
bound values based on numerical investigations. To this end, we first outline
the numerical method, then show results for $m=2$ and also beyond $m=2$ up
to $m=6$. In section~\ref{family}, a family of Bell inequalities
is presented (valid for any number of parties $N$), which allows
us to obtain good lower bounds for $P_{NL}^2(W_N)$. The paper ends
with a discussion in section~\ref{disc}.

\section{Bell setup}
\label{setup}

Let us imagine the following Bell setup~\cite{bell}. A quantum
state $\rho$ is shared between $n$ spatially separated systems, on
which the local observers can conduct measurements. We focus
on binary outcome measurements in which case we may define the
joint correlators by the following set of expectation values:
\begin{equation}\label{correlators}
\left\{\langle \mathcal{M}_{j_1}^{(1)}\ldots \mathcal{M}^{(n)}_{j_n}\rangle\right\}=\left\{\Tr{\left(\rho \mathcal{M}_{j_1}^{(1)}\otimes \ldots\otimes \mathcal{M}_{j_n}^{(n)}\right)}\right\}
\end{equation}
with $j_l=0,1,\ldots,m$ and $l=1,\ldots,n$. We identify
$\mathcal{M}_0^{(l)}=\one$ and $\mathcal{M}_{j_l}^{(l)}$ refers to
the $j_l$th $\pm 1$-valued observable of party $l$. The
corresponding real vector of correlators defines a point in the
$(1+m)^n$ dimensional space of probabilities. Each member of the
set~(\ref{correlators}) has an order, which is given by the amount
of parties $o$ involving a non-trivial observable (i.e. not
involving $\mathcal{M}_0^{(l)}=\one$). In particular, those with
$o=n$ are usually called full-correlators, while those with $o=1$
are called one-body correlators or marginal terms.

A multipartite (2-outcome) Bell inequality~\cite{bell} is a linear
function of the above correlators~(\ref{correlators}),
\begin{equation}\label{bellineq}
\sum_{j_1=0}^{m}\ldots\sum_{j_n=0}^{m}{\alpha_{j_1,\ldots,j_n}}\langle
\mathcal{M}_{j_1}^{(1)}\ldots
\mathcal{M}^{(n)}_{j_n}\rangle\le\beta_c,
\end{equation}

\noindent where we denote by $\beta_c$ the bound which holds for any local
hidden variable model. These are the correlations which the
parties can simulate by merely using local strategies and some
shared classical information. The local correlations attainable this
way forms a polytope, the so-called Bell polytope, whose extremal
points consist of those vectors in (\ref{bellineq}) in which all
correlators factorize, that is,
\begin{equation}
\label{factor}
\langle\mathcal{M}_{j_1}^{(1)}\ldots\mathcal{M}^{(n)}_{j_n}\rangle=\langle\mathcal{M}_{j_1}^{(1)}\rangle\cdot\ldots\cdot\langle\mathcal{M}_{j_n}^{(n)}\rangle
\end{equation}
and the mean value of each single party
$\langle\mathcal{M}_{j_l}^{(l)}\rangle$, $(l=1,\ldots,n, j_l>0)$ equals
either $-1$ or $+1$. Let us define the persistency of nonlocality
of a multipartite state $\rho$ according to
Ref.~\cite{persistency} as follows. Let us take the partial trace
over $0<k<N$ systems $i_1,...,i_k\in \{1,...,N\}$, and denote the
$n=N-k$-party reduced state by $\rho_{red}$. The persistency of
nonlocality of $\rho$, $P_{NL}(\rho)$, is defined as the minimal
$k$ such that the reduced state $ \rho_{red}$ becomes local for at
least one set of subsystems $i_1,...,i_k$. In other words, the
correlators~(\ref{correlators}) obtained from local measurements
$\mathcal{M}_{j_l}^{(l)}$ on $\rho_{red}$ do not violate any
Bell inequality.

If we allow at most $m$ different measurement settings in the
Bell expression~(\ref{bellineq}), we arrive at $P^m_{NL}(\rho)$
which provides a lower bound to $P_{NL}(\rho)$ and for
$m\rightarrow\infty$ recovers $P_{NL}(\rho)$.

In this paper, we focus on computing $P^m_{NL}(\rho)$ for the
noiseless $N$-qubit W state \cite{Wstate}, $\rho = |W_N\rangle\langle W_N|$, where
\begin{equation}
|W_N\rangle=\frac{1}{\sqrt N}\left(|0\ldots 01\rangle + |0\ldots 10\rangle + \ldots + |10\ldots 0\rangle\right).
\end{equation}

We may consider this state as a state of an atomic ensemble, and
we assume that $k$ particles are lost from this ensemble. In that
case, the reduced state contains $n=N-k$ particles, and the density
matrix reads
\begin{equation}
\label{rhoNk}
\rho(N,k)=\frac{n}{N}|W_{n}\rangle\langle W_{n}| + \frac{k}{N}|0^{\otimes n}\rangle\langle 0^{\otimes n}|.
\end{equation}
Since the W state is permutationally symmetric, the reduced state
$\rho(N,k)$ does not depend on the particular set of subsystems
removed, which simplifies considerably the analysis of $P_{NL}^m(W_N)$.

In the next section we provide an upper bound of $N/2$ for
$P_{NL}^{m=2}(W_N)$ in case of arbitrary even $N$ number of
parties, whereas in Sec.~\ref{LB} we bound this quantity by $2N/5$ from below.

\section{An upper bound for the persistency of nonlocality of the W state}
\label{UB}

We first prove the following lemma:
\begin{lemma} \label{Lbound}
Let us have a $2n$-qudit permutationally invariant state,
$\rho\in(\C^d)^{\otimes 2n}$, where $d\ge 2$ and $n\ge 1$. By
tracing out any $n$ qudits, the remaining $n$-qudit system
$\rho_{red}\in(\C^d)^{\otimes n}$ cannot violate any two-setting
$n$-party Bell inequality with arbitrary number of outcomes.
\end{lemma}

\begin{proof}
We take $N=2n$ and denote the $n$-qudit reduced state of
an $N$-qudit permutationally invariant state $\rho$ by
$\rho_{red}$ and the two measurements conducted on
station $l\in\{1,\ldots,n\}$ by $M^{(l)}_{a_l|j_l}$, where $j_l=1,2$. By permutationally invariance we mean that the exchange of any two qudits of the $N$-qudit state $\rho$ does not change the state itself. Then, the $n$-particle joint probability distribution reads
\begin{align}
\label{half}
&P(a_1,a_2,...,a_n|j_1,j_2,...,j_n)\nonumber\\
&=\Tr\left(\rho_{red}\mathcal{M}^{(1)}_{a_1|j_1}\otimes
\mathcal{M}^{(2)}_{a_2|j_2}\ldots\otimes
\mathcal{M}^{(n)}_{a_n|j_n}\right),
\end{align}
It is not difficult to see that the same probability distribution
can be achieved in the following way: Let us redistribute the
permutationally invariant state $\rho$ between $n$ parties such
that the $l$th qudit pair $(l,l+n)$ belongs to party
$l\in\{1,\ldots,n\}$ (so that each party owns two qudits). Then
party $l$ performs measurement $\mathcal{M}^{(l)}_{a_l|j_l=1}$ on
the first qudit and measurement $\mathcal{M}^{(l)}_{a_l|j_l=2}$
on the second qudit of the $l$th pair. This generates the same
distribution as of Eq.~(\ref{half}). Since for any $l$ these two
measurements act on different subspaces, they are commuting. However, Bell inequality violation is not possible (with any number of outcomes) if the two alternative
measurements for each party are pairwise commuting \cite{terhal}.
\end{proof}

Since the W state is a permutationally invariant multiqubit state
(with local dimension $d=2$), Lemma~1 above directly applies to
our situation, hence we get the upper bound $P_{NL}^m(W_N)\le n$
for $m=2$ with $N=2n$. In other words, $P_{NL}^2(W_N)\le N/2$ for
even $N$.

Some notes are in order. (i) It is straightforward to extend
Lemma~1 to more than two settings as well. For multiple settings,
we get the general upper bound $P_{NL}^m(W_N)\le (m-1)n$ with
$N=mn$. However, we conjecture that these upper bounds are not
tight in general. For $m=2$, a gap between the lower and upper
bound values for the $6\le N\le 20$-qubit W states are supported
by a numerical study performed in Sec.~\ref{LB}. For $m$ infinite,
the trivial upper bound $P_{NL}(W_N)\le N-1$ follows by plugging
$n=1$ in the above formula. If this bound happened to be tight,
it would imply that the two-qubit reduced state
$\rho_{red}=(2/N)|\psi^+\rangle\langle\psi^+| +
(1-(2/N))|00\rangle\langle 00|$ of the $W_N$ state was Bell
nonlocal. For large $N$ this is very unlikely, since the weight of
the entangled part $|\psi^+\rangle\langle\psi^+|$ goes to zero. In
fact, a recent computer study in Ref.~\cite{nagy} suggests that
the state $\rho(p)=p|\psi^+\rangle\langle\psi^+| +
(1-p)|00\rangle\langle 00|$ is local for $p<1/\sqrt 2$. Similarly,
A. Amirtham conjectures in Ref.~\cite{ami} that the state
$\rho(p)$ is local for $p<2/3$.

(ii) The permutational invariance property of the state is crucial in the
above state. If the multipartite state does not possess this high
symmetry, e.g. it only obeys translational invariance, the above
theorem does not hold true any more. Let us illustrate this with a
simple example. We consider a 4-qubit translationally invariant
state $\ket{\psi}=\ket{\psi^+}_{13}\ket{\psi^+}_{24}$, where
$\ket{\psi^+}_{ij}=(\ket{0}_i\ket{1}_j+\ket{1}_i\ket{0}_j)/\sqrt
2$.  Let us trace out particles $2$ and $4$ (constituting half of
the 4 particles), and as a result we get a maximally entangled
pair of qubits, which violates maximally the bipartite CHSH-Bell
inequality~\cite{chsh}. Nevertheless, translationally invariant
systems also impose certain restrictions which can be exploited in
Bell scenarios as studied in Ref.~\cite{Oliveira}.

\section{A lower bound for the persistency of nonlocality of the W state}
\label{LB}

We now give a numerical procedure which allows us to get useful
(and often tight) lower bounds to $P_{NL}^m(W_N)$. We note that
this procedure with some modification can also be applied to
generic permutationally invariant multiqubit states.

We consider the Bell violation of the following one-parameter
family of states:
\begin{equation}
\label{rhonp} \rho(n,p)=(1-p)|W_{n}\rangle\langle W_{n}| +
p|0^{\otimes n}\rangle\langle 0^{\otimes n}|.
\end{equation}
Notice that the state~(\ref{rhonp}) reproduces (\ref{rhoNk}) with
$p=k/N$. Hence, given an $n$-party $m$-setting Bell inequality
which is violated by state~(\ref{rhonp}) with a given critical
value, $p_{crit}$, we get the following lower bound on the
persistency:
\begin{equation}
\label{P_NL}
 P_{NL}^m(W_N)=N-n+1,
\end{equation}
where $N=\lfloor n/(1-p_{crit})\rfloor$, where $\lfloor x\rfloor$
maps a real number $x$ to the largest previous integer. Hence, in
order to get good lower bounds to $P_{NL}^m(W_N)$ our task reduces
to get good upper bounds to $p_{crit}$. To this end, we introduce
the following linear programming based numerical method.

\begin{figure}[htb]
\begin{center}
\label{figgeom}
\includegraphics[trim=2cm 6cm 9cm 3cm,clip=true,width=\columnwidth]{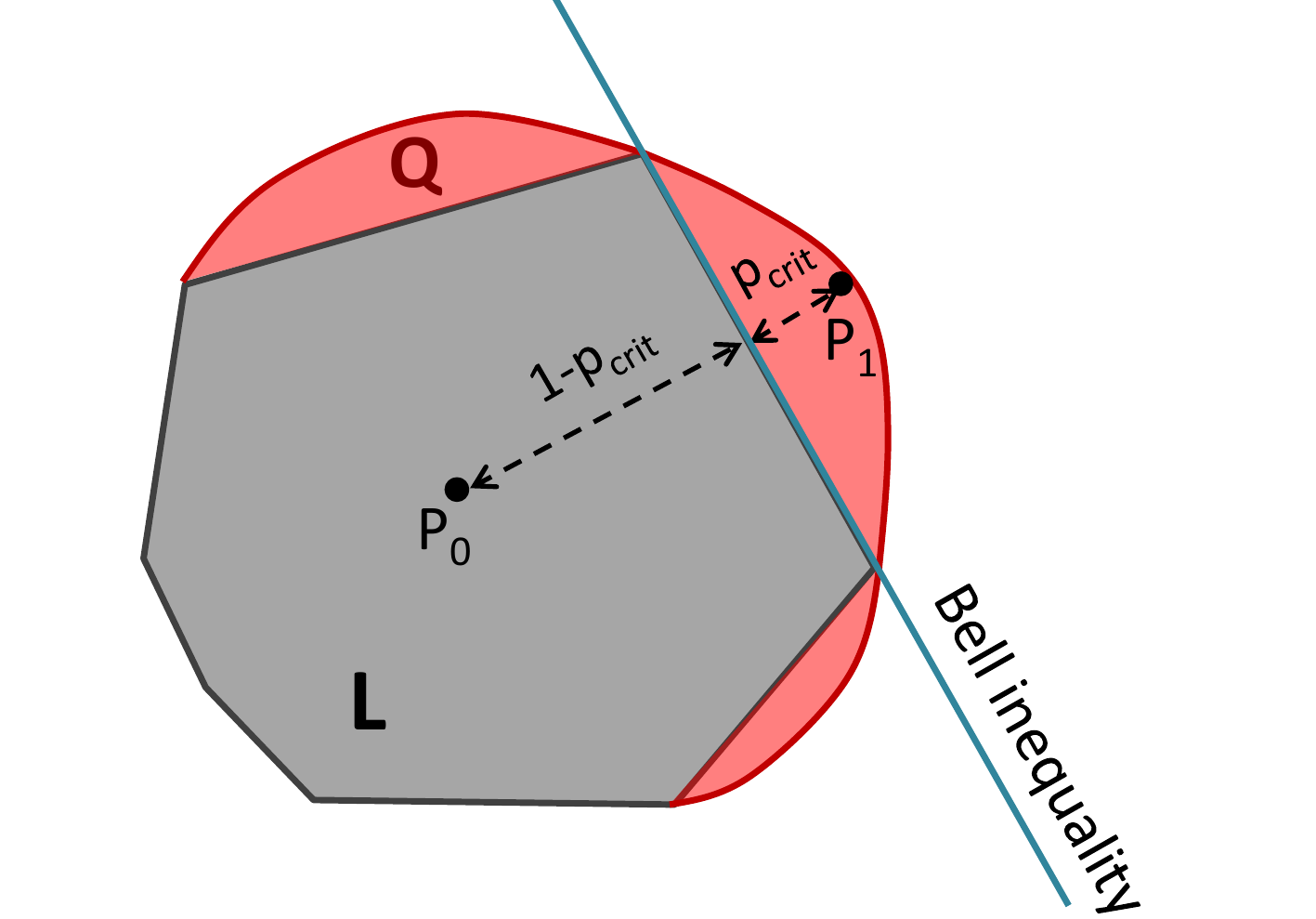}
\caption{A schematic view of the Bell polytope (region L) and the
space attainable with quantum systems (region Q). The correlation
point $P_1$ corresponding to the $W_N$ state is outside the Bell
polytope, whereas the point $P_0$ corresponding to the product
state falls inside the Bell polytope. $p_{crit}$ is fully
determined by points $P_0,P_1$ and the facets of the polytope.}
\label{Bellpolytope}
\end{center}
\end{figure}

Let us stick to $m=2$ settings (generalization to more settings is
straightforward). For simplicity we assume that all parties
measure the same qubit observables, that is,
$\mathcal{M}^{(1)}_j=\mathcal{M}^{(2)}_j=\ldots=\mathcal{M}^{(n)}_j$
for $j=1,2$. Moreover, due to symmetry of the states, we assume
these observables are coplanar, lying on the $X-Z$ equatorial
plane. Other works (e.g., Refs.~\cite{otherW,structure,tomoW})
maximizing Bell functionals using the W state rely on the same
symmetry considerations. With this simplification, we have two
optimization parameters. Hence, for the state $W_n$ and the above
measurements, the $(1+m)^n=3^n$-dimensional correlation point
$P_1$ given by the set of correlators~(\ref{correlators}) is
defined by two angles. Likewise, we define the correlation point
$P_0$ generated by the same measurements and the state
$\rho=|0^{\otimes n}\rangle\langle 0^{\otimes n}|$ in
Eq.~(\ref{correlators}). Geometrically, the correlations
accessible in a local hidden variables theory form a polytope, the
so-called Bell polytope, with vertices defined by deterministic
classical strategies for a fixed scenario of $n$ parties and $m$
settings (see Ref.~\cite{bellreview} for a review). The two
correlation points $P_0$ and $P_1$ are situated within this space.
Since the product state $|0^{\otimes n}\rangle\langle 0^{\otimes
n}|$ is local, point $P_0$ sits inside the Bell polytope, whereas
point $P_1$ depends on the two measurement angles and may well
fall outside the Bell polytope (see Fig.~\ref{Bellpolytope}). For
a given $p$ in (\ref{rhonp}), the corresponding correlation point
is $(1-p)P_1+pP_0$, and $p_{crit}$ is given by the intersection of
the line joining points $P_0$ and $P_1$ with the boundary of the
polytope (see Figure~\ref{Bellpolytope}). Given the two
measurement angles, standard linear programming allows us to
compute $p_{crit}$ and the underlying facet, which corresponds to
a Bell inequality.

\begin{figure}[htb]
\begin{center}
\includegraphics[width =\columnwidth]{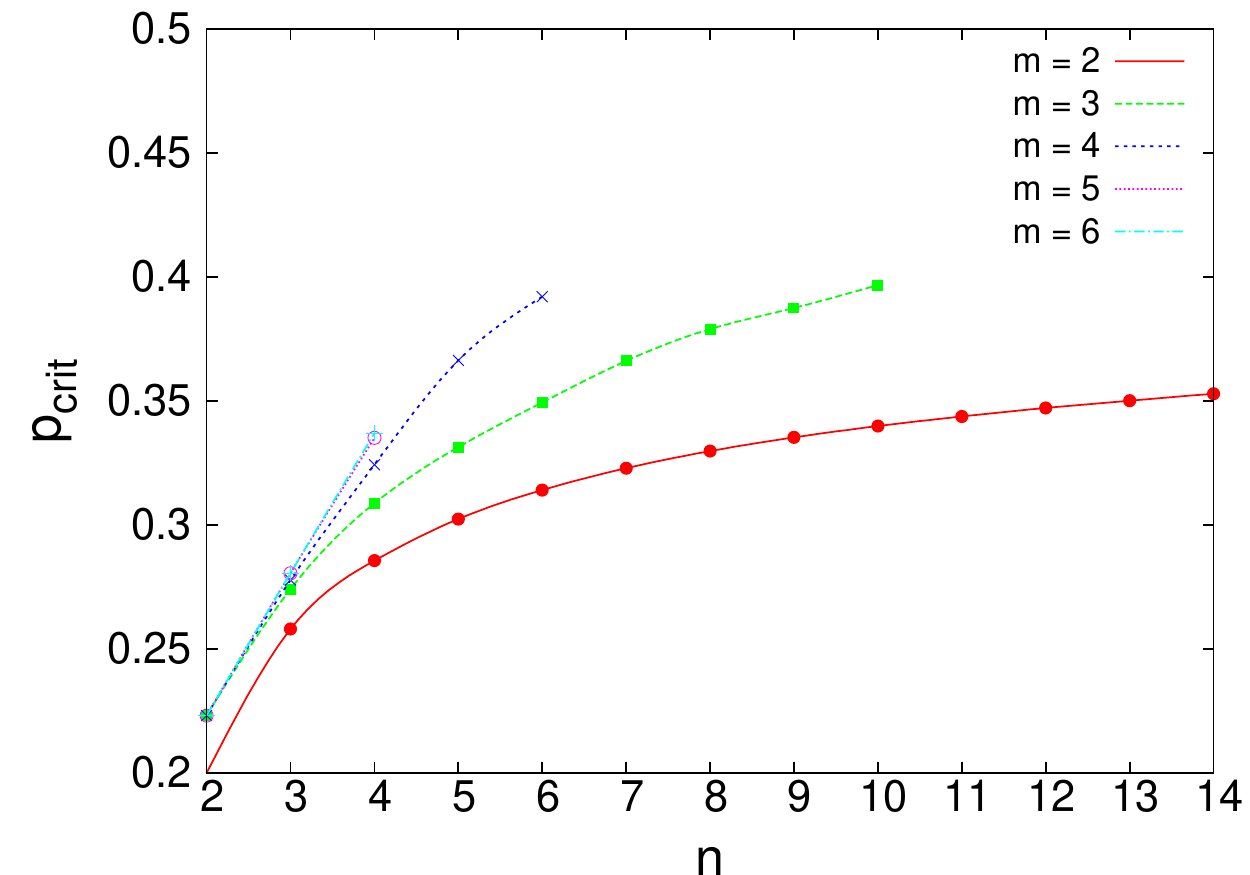}
\hspace{2cm}
\caption{For $\rho(n,p)$ states of $n$ qubits (\ref{rhonp}), the graph shows lower bounds on the threshold
probability $p_{crit}$ obtained using our numerical method up to
$m=6$ measurement settings.} \label{Wm6}
\end{center}
\end{figure}

Clearly, the above described procedure works for $m>2$ as well. In
particular, we have chosen a given Bell scenario ($n$ parties, $m$
settings) and by varying the $m$ angles, we maximized the value of
$p_{crit}$. We note that this search is a heuristic one and as
such it is not guaranteed to terminate in a global maximum of
$p_{crit}$. However, the obtained value still defines a lower
bound to $P^m_{NL}$ in (\ref{P_NL}). The critical $p$ values
obtained in function of $n$ and $m$ are displayed in
Fig.~\ref{Wm6}. We may observe that, as $m$ increases, the
affordable number of parties $n$ decreases. For the simplest case
of $m=2$, numerically we could afford $n=14$. In addition, we had
to resort to a symmetrization procedure of the Bell polytope
introduced in \cite{JD} which considerably reduces the complexity
of the problem in order to attain $n=14$.

Plugging the $p_{crit}$ values shown in Fig.~\ref{Wm6} into
formula~(\ref{P_NL}), we get Table~\ref{tabla}, where the computed
persistencies $P_{NL}^m$ are shown for $m\le 6$. This way we get
entries in the table only for $N$ satisfying $N=\lfloor
n/(1-p_{crit})\rfloor$, but a slight modification allows us to
obtain numbers for any $N$ parties: Fix $N$, and choose the
largest $k$ integer such that $p_{crit,n-k}>k/N$, where
$p_{crit,n-k}$ stands for $p_{crit}$ evaluated by the number of
parties $n-k$. Then a lower bound on the persistency $P_{NL}^m$
for the $N$-qubit W state is given by $P_{NL}^m=k+1$.

\begin{table} \begin{tabular}{c|ccccc}
\hline \hline
     N & $m=2$ & $m=3$ & $m=4$ & $m=5$ & $m=6$ \\
     \hline
     2  &   1  &   1  &   1  &   1  &   1\\
     3  &   1  &   1  &   1  &   1  &   1\\
     4  &   2  &   2  &   2  &   2  &   2\\
     5  &   2  &   2  &   2  &   2  &   2\\
     6  &   2  &   2  &   2  &   3  &   3\\
     7  &   3  &   3  &   3  &      &    \\
     8  &   3  &   3  &   3  &      &    \\
     9  &   3  &   4  &   4  &      &    \\
    10  &   4  &   4  &      &      &    \\
    11  &   4  &   5  &      &      &    \\
    12  &   4  &   5  &      &      &    \\
    13  &   5  &   5  &      &      &    \\
    14  &   5  &   6  &      &      &    \\
    15  &   6  &   6  &      &      &    \\
    16  &   6  &   7  &      &      &    \\
    17  &   6  &      &      &      &    \\
    18  &   7  &      &      &      &    \\
    19  &   7  &      &      &      &    \\
    20  &   8  &      &      &      &    \\
\hline\hline
\end{tabular}
\caption{\label{tabla} Lower bound values for persistency of
nonlocality $P_{NL}^m$ for the $W_N$ state with $m=2,\ldots,6$
binary-outcome measurement settings per party. The results are
based on the numerical method described in the text.}
\end{table}

\begin{figure}[htb]
\begin{center}
\includegraphics[width =\columnwidth]{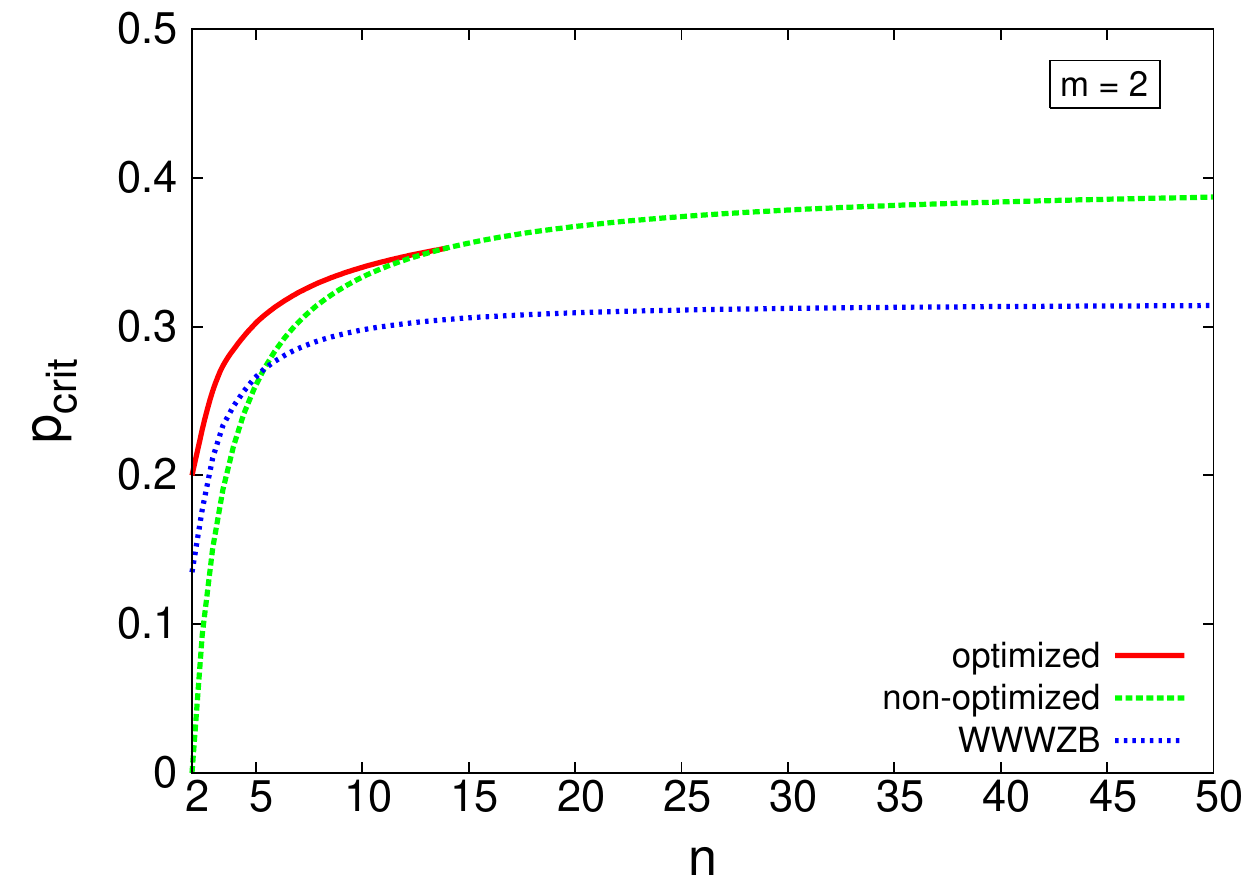}
\caption{For $\rho(n,p)$ states of $n$ qubits (\ref{rhonp}), the
graph shows lower bounds on the threshold probability $p_{crit}$
obtained using our numerical method for $m=2$ measurement
settings. These bounds coincide with the bounds coming from our
class of Bell inequalities for 2 measurement settings up to $n=14$
and we conjecture to be the same for any even $n$ particles. We
show results for our Bell inequalities with optimized and
non-optimized settings (in the latter case, the two observables
are the Pauli operators $Z$ and $X$). The lower curve reproduces
the results of Ref.~\cite{tomer} in case of the WWWZB
inequalities~\cite{WWWZB}.} \label{Wm2}
\end{center}
\end{figure}

In Table~\ref{tabla}, for $m=2$, the first entries
$N=2,3,\ldots,7$ match the numbers from previous study
\cite{persistency}. However, by fixing $N$ and going to higher
number of settings $m$ sometimes we get a higher persistency of
nonlocality. For instance, in case of $N=6$ parties, $P_{NL}^5=3$,
which is to be compared to $P_{NL}^2=2$. For $m=2$, we extracted
the optimal Bell inequalities corresponding to $p_{crit}$ for
various $n$ values. For $n$ even, a common structure has been
found. In fact, they turned out to be members of a family of Bell
inequalities valid for any $n$ even. Details of this class of
inequalities are presented in Sec.~\ref{family}. We optimized the
quantum value of these inequalities in function of the two
measurement angles. Fig.~\ref{Wm2} shows results for $p_{crit}$ up
to $n=50$ of this family. We also show results using the ansatz
that the two measurement settings are $Z$ and $X$, in which case
$p_{crit}=(2n-4)/(5n-2)$ for $n$ even. According to the figure,
for larger $n$ the $(Z,X)$ pair of settings become close to
optimal. The $p_{crit}$ values have been also studied in
Ref.~\cite{tomer} for specific families of known Bell inequalities
from the literature, with the best lower bound values coming from
the WWWZB inequalities~\cite{WWWZB}. For comparison, these
$p_{crit}$ values are displayed. For $n=50$, $p_{crit}=0.3142$,
which is suboptimal with respect to our
$p_{crit}=(2n-4)/(5n-2)\simeq0.3871$.

\begin{figure}[htb]
\begin{center}
\includegraphics[width =\columnwidth]{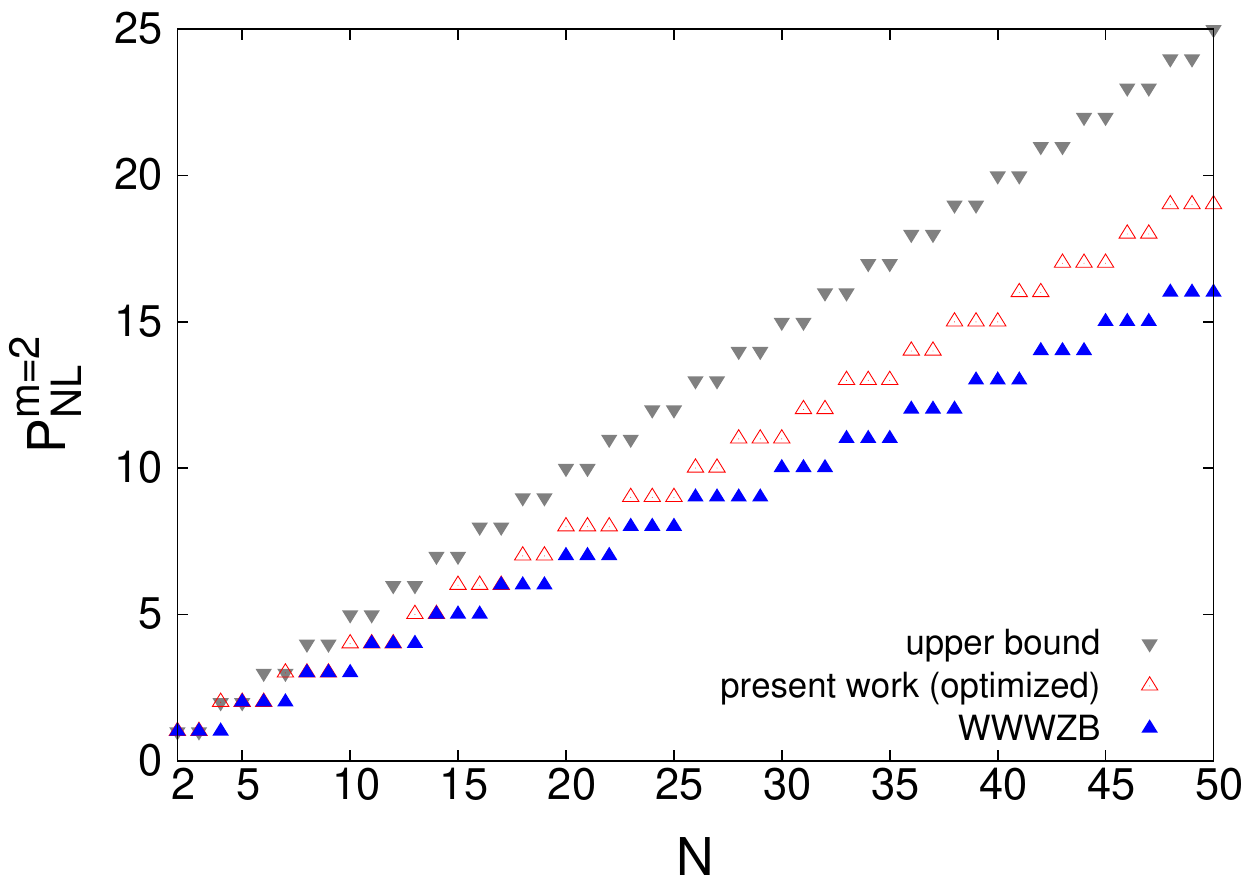}
\caption{For $W$ states of $N$ qubits, the graph shows upper
bounds (marked by a triangle pointing down) and lower bounds (marked by triangles pointing up) on the persistency of nonlocality
$P_{NL}^{m=2}(W_N)$.} \label{Wm2persistency}
\end{center}
\end{figure}

Finally, Fig.~\ref{Wm2persistency} shows the upper bound values
due to section~\ref{UB} and the lower bound values for
$P^2_{NL}(W_N)$ up to $N=50$ due to our Bell family. We conjecture
that the exact value for $P^2_{NL}(W_N)$ is defined by the lower
curve (that is, no better Bell inequalities than our family in
section~\ref{family} exist in this respect). For instance, the
best lower bound for $P^2_{NL}(W_N)$ so far comes from the WWWZB
inequalities analyzed in Ref.~\cite{tomer}, which lies below our
curve for the lower bound, and according to the conjecture in
Ref.~\cite{tomer} it goes to $N/3$ in the limit of large $N$. On
the other hand, our asymptotic lower bound value of $2N/5$ is
considerable higher. It is also worth noting that the upper bound
value on $P^2_{NL}$ shown in the figure holds true for any
permutationally invariant $N$-qubit state. Hence, Dicke states
with more than one excitation, as a notable subset of
permutationally invariant states, cannot perform significantly
better than $W$ states in the two-setting case.

In Ref.~\cite{tomer}, the Dicke states (including the $W_N$ state)
were analyzed in terms of two decoherence models: loss of
particles and loss of excitations. The first one relates to the
measure $P_{NL}$ discussed in this paper, whereas the latter one
is related to $p_{crit}$ in formula~(\ref{rhonp}). Indeed, if we
start from an $n$-qubit W state which is effected by a decoherence
where in each mode an excitation has a probability $p$ of being
lost, it brings the W state into (\ref{rhonp}). Appendix~A shows a
derivation of this result.

Next section gives all the details, including the quantum and
local bounds, of our particular family of two-setting Bell
inequalities providing $p_{crit}=(2n-4)/(5n-2)$ for even $n$. In
case of large $n$, this goes to $2/5$ which we conjecture to be
the largest achievable critical value among all two-setting Bell
inequalities.

\section{A family of two-setting multipartite Bell inequalities}
\label{family}

\textit{Standard form of permutationally symmetric Bell inequalities.--} The Bell inequality to be considered consists of correlators~(\ref{correlators}) which are invariant under any permutation of the parties. By imposing the permutational symmetry, one requires that the expectation values $\langle\mathcal{M}_{j_1}^{(1)}\ldots
\mathcal{M}^{(n)}_{j_n}\rangle$ and $\langle\mathcal{M}_{j_1}^{\sigma(1)}\ldots
\mathcal{M}^{\sigma(n)}_{j_n}\rangle$ are the same, where $\sigma:\{1,\ldots,n\}\rightarrow\{\sigma(1),\ldots,\sigma(n)\}$ is an arbitrary permutation of the set $\{1,\ldots,n\}$. Let us denote by $S_n$ all permutations of the set $\{1,\ldots,n\}$.
Then, it will be useful to define
\begin{equation}\label{Sor}
S^o_r \equiv \sum_{\sigma\in S_n}\langle\mathcal{M}_{j_1}^{\sigma(1)}\ldots
\mathcal{M}^{\sigma(n)}_{j_n}\rangle,
\end{equation}

\noindent where the sum ranges over all permutations of the set $\{1,\ldots,n\}$. In $S^o_r$ above, $o$ is the order of correlators and $r$ is the number of $1$'s occurring in the list $\{j_1,\ldots,j_n\}$. Let us now define the symmetrized correlation vector by the above ordered real vectors as follows
\begin{equation}
\vec S \equiv \{S^o_r\}^{o=1,\ldots,n}_{r=0,\ldots,o}.
\end{equation}

\noindent In a similar way, we define the vector of coefficients associated with $\vec S$ as
\begin{equation}
\vec \alpha \equiv \{\alpha^o_r\}^{o=1,\ldots,n}_{r=0,\ldots,o}.
\end{equation}

\noindent As a result, we arrive at the general form of a permutationally symmetric Bell inequality:
\begin{equation}
\label{symBI}
I \equiv \vec\alpha\cdot\vec S = \sum_{\substack{{o=1,\ldots,n}\\{r=0,\ldots,o}}}{\alpha^o_r S^o_r}\le\beta_c,
\end{equation}
where $\beta_c$ is the local maximum, which holds for any local
hidden variable model. It is worth noting that the WWWZB
class~\cite{WWWZB} contains only full-correlator terms ($o=n$),
whereas our construction of Bell inequalities turns out to contain
all different orders, starting from marginal terms ($o=1$) up to
full-correlation terms ($o=n$). There also exist permutationally
symmetric $n$-party Bell inequalities involving only first and
second order terms ($o=1,2$)~\cite{twobody}. The usefulness of
these kind of inequalities in the particle loss model
is an open and interesting question in our view.

As an illustrative example, let us discuss the case of 4 particles
($n=4$). In this case, $\vec S$ looks as follows
\begin{equation}
\vec S = \{S^1_1,S^1_0;S^2_2,S^2_1,S^2_0;S^3_3,S^3_2,S^3_1,S^3_0;S^4_4,S^4_3,S^4_2,S^4_1,S^4_0\},
\end{equation}
where we used semicolon (;) in order to separate components with
different order $o$. By identifying $\mathcal{M}_0^{(l)}=1$, $l=1,\ldots,4$ and
$\mathcal{M}_j^{(1)}=A_j$, $\mathcal{M}_j^{(2)}=B_j$,
$\mathcal{M}_j^{(3)}=C_j$, $\mathcal{M}_j^{(4)}=D_j$ for $j=1,2$
in Eq.~(\ref{Sor}), we have the following mean values displayed
for some particular cases (neglecting the mean value signs):
\begin{align}
\label{Sexample}
S^1_1 =& 6(A_1 + B_1 + C_1 + D_1),\nonumber\\
S^2_2 =& 4(A_1B_1 + A_1C_1 + A_1D_1 + B_1C_1 + B_1D_1 + C_1D_1),\nonumber\\
S^2_0 =& 4(A_2B_2 + A_2C_2 + A_2D_2 + B_2C_2 + B_2D_2 + C_2D_2),\nonumber\\
S^3_1 =& 2(A_1B_2C_2+A_1B_2D_2+A_1C_2D_2)\nonumber\\
& +2(A_2B_1C_2+A_2B_1D_2+B_1C_2D_2)\nonumber\\
& +2(A_2B_2C_1+A_2C_1D_2+B_2C_1D_2)\nonumber\\
& + 2(A_2B_2D_1+A_2C_2D_1+B_2C_2D_1),\nonumber\\
S^4_0 =& 24A_2B_2C_2D_2,
\end{align}

\noindent where we recall that $o$ denotes the order in the
superindex of $S^o_r$, whereas $r$ is the number of occurrences of
subindex 1 in $S^o_r$. For instance, $S^3_1$ above consists of all
length-3 sequences from $A,B,C$ and $D$ letters with the occurrence of a single
subindex 1.

\textit{The specific class of Bell inequalities.--} Let us now
give the explicit form of a Bell inequality $I$ defined by the
vector of coefficients $\vec\alpha$ in Eq.~(\ref{symBI}). This
family has been extracted from the numerical method of
Sec.~\ref{LB}. In particular, we introduce a family of Bell
inequalities $I_n$ valid for any even number of particles $n$ as
follows
\begin{equation}
\label{nsymBI}
I_n \equiv \vec\alpha_n\cdot\vec S_n = \sum_{\substack{{o=1,\ldots,n}\\{r=0,\ldots,o}}}{\alpha^o_r S^o_r}\le\beta_{c,n}.
\end{equation}

\noindent The vector of coefficients $\vec\alpha=\{\alpha^o_r
(o=1,\ldots,n, r=0,\ldots,o)\}$ reads as

\begin{equation}\label{nalpha}
\vec\alpha=
\begin{cases}
    \alpha^{2k}_0 = F_{k,n}, & k=1,\ldots,n/2\\
    \alpha^{2k-1}_1 = G_{k,n}+2w_n\delta_{k,1}, & k=1,\ldots,n/2\\
    \alpha^2_2 = -w_n,\\
    \alpha^o_r = 0, & \text{otherwise}
\end{cases}
\end{equation}
where $\delta_{i,j}$ is the Kronecker delta (1 if $i=j$, 0
otherwise), and the functions entering eq.~(\ref{nalpha}) above
are
\begin{align}
F_{k,n}&=(-1)^k\frac{n-2}{2}(n+1-2k)\binom{n/2}{k},\nonumber\\
G_{k,n}&=(-1)^k\frac{n+2}{2}n\binom{n/2-1}{k-1},\nonumber\\
w_n&=\frac{n(n-1)(n+2)}{2^{4-n}\binom{n}{n/2}}.
\end{align}

\noindent Especially for $n=4$, our 4-party Bell inequality looks
as follows:
\begin{equation}
I_4 = 12 S^1_1 - 12 S^2_2 - 6S^2_0 + 12S^3_1 + S^4_0 \le 128,
\end{equation}
\noindent where the middle terms appear explicitly
in~(\ref{Sexample}) and the local bound is due to
formula~(\ref{betacn}) given in the next subsection for the
general $n$-party case.

\textit{Local bound.--} The local maximum, which is the maximum value one obtains using local resources only, is given by
\begin{equation}
\label{betacn}
\beta_{c,n}=n!\left(w_n-\frac{(n-2)(n+1)}{2}\right)
\end{equation}
in case of any even number of particles $n$, where
\begin{equation}
w_n=\frac{n(n-1)(n+2)}{2^{4-n}\binom{n}{n/2}}.
\end{equation}

\noindent We have checked the validity of the above local bound
$\beta_{c,n}$ for  $n\le 16$ ($n$ even) by listing all possible
deterministic strategies (i.e., vertices of the Bell polytope).
These are the strategies for which all correlators factorize (see
eq.~\ref{factor}) and each single party marginal
$\langle\mathcal{M}_{j_l}^{(l)}\rangle$, $(l=1,\ldots,n, j_l=1,2)$
equals either $-1$ or $+1$. One of the above deterministic
strategies, due to linearity of the Bell functional, will provide the
local maximum. Notice that $\langle\mathcal{M}_0^{(l)}\rangle$,
($l=1,\ldots,n$) equals $+1$ by definition. Below we provide an
analytical proof of formula~(\ref{betacn}) for any even $n$. We
checked with brute-force computations that for smaller values of
$n$ the inequalities are far from tightness, that is, they do not
define a facet of the Bell polytope. We conjecture that they are
not tight for larger $n$ as well.

Here follows the proof of the local limit~(\ref{betacn}) for even
$n$. First let us fix notation. For a given local deterministic
strategy let us define a four-tuple ($a$,$b$,$c$,$d$) which counts
the number of parties whose marginal expectations
$\{\langle\mathcal{M}^{(l)}_1\rangle,\{\langle\mathcal{M}^{(l)}_2\rangle\}$ are the respective
pairs ($\{1,1\}$, $\{1,-1\}$, $\{-1,1\}$, $\{-1,-1\}$). By
definition we have $a+b+c+d=n$. In case of permutationally
invariance under party exchange, a four-tuple ($a,b,c,d$)
represents uniquely a deterministic strategy. Hence, our task is
to calculate the Bell value $\vec\alpha_n\cdot\vec S_n$ for all
possible integer-valued four-tuples $a,b,c,d\ge 0$ fulfilling
$a+b+c+d=n$, and then pick the maximum value out of this set. This
defines the local maximum $\beta_{c,n}$ attainable using classical
resources. In the proof below, all positive integer four-tuples
$a,b,c,d$ are understood to sum up to $n$, where $n$ is even.

Let us introduce the permanent of an $n\times n$ matrix $A=(a_{i,j})$, $(i,j=1,\ldots,n)$. It is defined as
\begin{equation}
\text{perm}(A) = \sum_{\sigma\in S_n}\prod_{i=1}^n{a_{i,\sigma(i)}},
\end{equation}
where $\sigma$ is a permutation over the set $\{1,\ldots,n\}$. Using the above definition for the permanent, $\vec S = \{S^o_r\}$ ($o=1,\ldots,n$, $r=0,\ldots,o$) is given by the components $S^o_r=\text{perm}(M^{o,r})$ for a given deterministic strategy $a,b,c,d$, where $M^{o,r}(a,b,c,d)$ is an $n\times n$, $\pm 1$-valued matrix whose components $u,v=1,\ldots,n$ are defined as follows


\begin{equation}
M^{o,r}_{u,v}(a,b,c,d)=\nonumber
\end{equation}

\begin{equation}
\begin{cases}
    -1,& \text{if } \quad n-o<u\le n-o+r\quad\text{and }\quad a+b<v\\
    -1,& \text{if } \quad n-o+r< u\quad\text{and }\\
       &\quad\quad (a<v\le a+b\quad \text{or }\quad a+b+c<v)\\
    +1,              & \text{otherwise.}
\end{cases}
\end{equation}
However, for $S^o_1=\text{perm}(M^{o,1})$, $o=1,\ldots,n$, there exists a closed form expression as well (which we will make use of later):
\begin{align}
&S^o_1 =(o-1)!(n-o)!\nonumber\\
&((a-c)\sum_{k=0}^n{(-1)^k\binom{n-a-c}{k}\binom{a+c-1}{o-1-k}}+(a+2b+c-n)\nonumber\\
&\sum_{k=0}^n{(-1)^k\binom{n-a-c-1}{k}\binom{a+c}{o-1-k}}).
\end{align}

\noindent This expression comes from an expansion of matrix $M^{o,r}$ in terms of row $(n-o)$. Let us now define two auxiliary functions $f_{1,2}$ which later will prove to be useful:
\begin{equation}\label{f1}
 f_1(n,b+d)=
\begin{cases}
    2,& \text{if } 2(b+d)=n\\
    1,& \text{if } |2(b+d)-n|=2\\
    0,              & \text{otherwise}
\end{cases}
\end{equation}
and
\begin{equation}\label{f2}
 f_2(a+b+c+d)=
\begin{cases}
    a+b-c-d,& \text{if } a+c=b+d\\
    a-c,& \text{if } a+c=b+d+2\\
    b-d,& \text{if } a+c=b+d-2\\
    0,              & \text{otherwise}
\end{cases}
\end{equation}
Recall that $a+b+c+d=n$. Let us also recall that the Bell expression for a particular deterministic strategy looks as follows
\begin{equation}
I(a,b,c,d)= \vec\alpha\cdot\vec S(a,b,c,d),
\end{equation}
where $\vec\alpha$ defines the Bell coefficients through eq.~(\ref{nalpha}). Let us divide the terms appearing in~(\ref{nalpha}) into three distinct cases $\vec\alpha =\vec\alpha_{I} + \vec\alpha_{II} + \vec\alpha_{III}$ as follows:

\begin{equation}\label{nalphaI}
\vec\alpha_{I}=
\begin{cases}
    \alpha^{2k-1}_{I,1} = 2w_n\delta_{k,1}, & k=1,\ldots,n/2\\
    \alpha^2_{I,2} = -w_n,\\
    \alpha^o_{I,r} = 0, & \text{otherwise},
\end{cases}
\end{equation}

\begin{equation}\label{nalphaII}
\vec\alpha_{II}=
\begin{cases}
    \alpha^{2k}_{II,0} = F_{k,n}, & k=1,\ldots,n/2\\
    \alpha^o_{II,r} = 0, & \text{otherwise},
\end{cases}
\end{equation}

\begin{equation}\label{nalphaIII}
\vec\alpha_{III}=
\begin{cases}
    \alpha^{2k-1}_{III,1} = G_{k,n}, & k=1,\ldots,n/2\\
    \alpha^o_{III,r} = 0, & \text{otherwise}.
\end{cases}
\end{equation}

\noindent Using the above formulas, one can show (after tedious but straightforward calculations) that

\begin{align}\label{threebi}
\vec\alpha_{I}\cdot\vec S(a,b,c,d) =& w_n(n-2)!\left(n(n-1)+4(c+d)(1-c-d)\right)\nonumber\\
\vec\alpha_{II}\cdot\vec S(a,b,c,d) =& \left(\frac{n}{2}-1\right)(2^{n-1}\left(\frac{n}{2}+1\right)\left(\frac{n}{2}!\right)^2 f_1(n,b+d)\nonumber\\
&-(n+1)!)\nonumber\\
\vec\alpha_{III}\cdot\vec S(a,b,c,d) =& -f_2(a,b,c,d)2^{n-1}\frac{n}{2}!\left(\frac{n}{2}+1\right)!
\end{align}

\noindent Furthermore, summing up the above three equations~(\ref{threebi}), we arrive at

\begin{align}
&I(a,b,c,d) = \vec\alpha\cdot\vec S(a,b,c,d) = \nonumber\\
&\left(\frac{n}{2}!\right)^2 2^{n-2}\left(\frac{n}{2}+1\right)2
\left(f_1(n,b+d)\left(\frac{n}{2}-1\right)-f_2(a,b,c,d)\right)\nonumber\\
&+ \left(\frac{n}{2}!\right)^2 2^{n-2}\left(\frac{n}{2}+1\right)\left(\frac{n}{2}(n-1)+2(c+d)(1-c-d)\right)\nonumber\\
&-(\frac{n}{2}-1)(n+1)!
\end{align}

\noindent Now we separate the above expression into four different cases according to the numbers $a,b,c,d$ occurring in the auxiliary functions $f_1$ and $f_2$ in eqs.~(\ref{f1},\ref{f2}). After a bit of manipulation, we arrive at

\begin{align}\label{bix}
&I(a,b,c,d)=\left(\frac{n}{2}!\right)^2 2^{n-2}\left(\frac{n}{2}+1\right)x\nonumber\\
&+ \left(\frac{n}{2}!\right)^2 2^{n-2}\left(\frac{n}{2}+1\right)\left(\frac{n}{2}(n-1)+2(c+d)(1-c-d)\right)\nonumber\\
&-(\frac{n}{2}-1)(n+1)!
\end{align}
where
\begin{equation}\label{x}
 x=
\begin{cases}
    -4+4(c+d),& \text{if } a+c=b+d\\
    -a+b+3c+d-2,& \text{if } a+c=b+d+2\\
    a-b+c+3d-2,& \text{if } a+c=b+d-2\\
    0,              & \text{otherwise.}
\end{cases}
\end{equation}
Let us subtract the conjectured $\beta_{c,n}$~(\ref{betacn}) from $I(a,b,c,d)$ in formula~(\ref{bix}). Notice that we end the proof once we find that $\beta_{c,n}-I(a,b,c,d)\ge 0$ for all possible positive integers $a,b,c,d$ with $a+b+c+d=n$.
Dividing by $\left(\frac{n}{2}!\right)^2 2^{n-2}$ and after some lengthy manipulation, we get

\begin{equation}\label{simple}
4\left(\frac{n}{2}+1\right)-\left(\frac{n}{2}+1\right)y\ge 0,
\end{equation}
with
\begin{equation}\label{y}
 y=
\begin{cases}
    4(c+d)+2(c+d)(1-c-d),& \text{if } a+c=b+d\\
    4c+2(c+d)(1-c-d),& \text{if } a+c=b+d+2\\
    4d+2(c+d)(1-c-d),& \text{if } a+c=b+d-2\\
    4+2(c+d)(1-c-d),              & \text{otherwise.}
\end{cases}
\end{equation}
It is straightforward to check that in each above case the maximum allowed value of $y$ is 4. Substituting this value back into~(\ref{simple}), we obtain that inequality~(\ref{simple}) is never violated. This completes the proof of the local bound $\beta_{c,n}$ expressed by formula~(\ref{betacn}).

\textit{Quantum violation.--} Hereby we give a closed form of the
vector $\vec S$ in case of qubit observables and an $n$-qubit
quantum state. In quantum theory, we have the expectation value
defined by Eq.~(\ref{correlators}). We specify the $n$-qubit state
$\rho$ to be the one-parameter family of states given by
$\rho(n,p)$ in eq.~(\ref{rhonp}). The qubit observables, on the
other hand, are chosen as $\mathcal{M}_1^{(l)}=Z$ and
$\mathcal{M}_2^{(l)}=X$ for all $l=1,\ldots,n$, where
$Z=|0\rangle\langle0|-|1\rangle\langle1|$ and
$X=|0\rangle\langle1|+|1\rangle\langle0|$ are the $2\times 2$ Pauli
matrices. Also, $\mathcal{M}_0^{(l)}=\one$ for all $l=1,\ldots,n$,
by definition.

Borrowing formulas from Ref.~\cite{structure}, we find a closed
form expression for $\vec S \equiv \{S^o_r\}$ ($o=1,\ldots,n$,
$r=0,\ldots,o$). Let us write
\begin{equation}
\label{vecS}
S^o_r=(1-p) P^o_r + p Q^o_r
\end{equation}
where we have
\begin{align}
Q^o_r =& n!\delta_{o,r},\nonumber\\
P^o_r =& (n-1)!\left((n-2r)\delta_{o,r} + 2\delta_{o,r+2}\right),
\end{align}
where $\delta_{i,j}$ stands for the Kronecker delta.

\textit{Critical value of $p$.--} Our next task is to compute the critical value $p_{crit}$ in function of $n$ for which we have $\vec\alpha\cdot\vec S = \beta_{c,n}$, where the components of $\vec S$ in Eq.~(\ref{vecS}) contain $p$ as a parameter. Note that $\vec\alpha$ and $\beta_{c,n}$ are defined through Eq.~(\ref{nalpha}) and Eq.~(\ref{betacn}), respectively. By substitution we arrive at
\begin{align}
\label{PQ}
\vec\alpha\cdot\vec P &= \sum_{\substack{{o=1,\ldots,n}\\{r=0,\ldots,o}}}{\alpha^o_r P^o_r}=n!\left(w_n-\frac{(n-2)(n-1)}{2}\right),\nonumber\\
\vec\alpha\cdot\vec Q &= \sum_{\substack{{o=1,\ldots,n}\\{r=0,\ldots,o}}}{\alpha^o_r Q^o_r}=n!\left(w_n-\frac{n(n+2)}{2}\right).
\end{align}

\noindent Next, applying the criterion $\vec\alpha\cdot\vec S = (1-p)\vec\alpha\cdot\vec P + p\vec\alpha\cdot\vec Q=\beta_{c,n}$, we have the critical value for $p$ as follows,
\begin{equation}
p_{crit}=\frac{\beta_{c,n}-\vec\alpha\cdot\vec P}{\vec\alpha\cdot\vec Q-\vec\alpha\cdot\vec P}=\frac{2n-4}{5n-2},
\end{equation}
where we used formulas~(\ref{PQ}) above.

Note that the formula for the critical value of $p=(2n-4)/(5n-2)$
above goes to $p_{crit}\rightarrow2/5$ as the number of particles
$n$ goes to infinity.

\section{Discussion}
\label{disc}

The multipartite W state is an important state relevant to the interaction between light and matter.
We addressed the persistency of the
nonlocality $P_{NL}$ of this state both by numerical and
analytical means. In case of two-setting measurements ($m=2$) we
could pin down the value of $P_{NL}^m$ such that there remains
only a relatively small gap between the upper and lower bound
values for any number of parties $N$. For $N$ large, this value
tends to be within the range $[2N/5,N/2]$. Moreover, based on a
numerical investigation regarding the lower bound value, we
conjecture that $2N/5$ is the exact value for large $N$. In this
respect, it would be interesting to improve further the upper
bound value. Note that the proof for the upper bound of
$P_{NL}^{2}$ in section~\ref{UB} relies merely on the
permutationally symmetry of the state and does not exploit the
full structure of the W state. On the other hand, our numerical study indicates that for a fixed but small $N$,
$P_{NL}^m(W_N)$ increases by increasing $m$. This suggests that for $N$ large the lower bound $2N/5$ on $P_{NL}^{m=2}(W_N)$ increases as well in case of $m>2$. Finding a general $N$-party $m$-setting family of Bell inequalities to
lowerbound $P_{NL}^m$ of which the present one is a special $m=2$
member would be most welcome.

Let us mention some possible ways to generalize the persistency of
nonlocality of multipartite states. The concept of EPR
steering~\cite{EPRsteering} lies between entanglement and
nonlocality, and EPR steering of multipartite quantum states has
been investigated recently~\cite{EPRmulti}. Similarly to the
persistency of nonlocality, it would be interesting to study the
behavior of persistency of steering for the W state or other
permutationally invariant states such as Dicke states. Finally,
the question of genuine nonlocality~\cite{svet} of the W state has
also been left open. Indeed, instead of studying the persistency
of standard nonlocality $P_{NL}$ of the W state, we may ask as
well what is the minimal number of parties $k$ to trace out from
an $N$-qubit W state, such that the reduced $N-k$-party state
lacks genuinely multipartite nonlocality.

\begin{acknowledgments}
We acknowledge financial support from the Hungarian National
Research Fund OTKA (K111734).
\end{acknowledgments}

\appendix

\section{Loss of excitations}
\label{appa}

Suppose that a source emits an $n$-qubit state $\rho$, which can be effected by some losses, e.g., noise due to a lossy channel.  We treat channel losses in the following noise model, which is called amplitude damping. In each
mode (out of $n$ modes), there is a probability $p$ of losing an excitation. The operator sum formalism (see e.g., Ref.~\cite{nielsenchuang}) describes the transformation between the initial state $\rho$ and the final state $\rho_{noisy}$ in the following way
\begin{equation}
\rho_{noisy}=\sum_{\vec{k}=(0,\dots,0)}^{(1,\dots,1)}\mathcal{K}_{\vec{k}}\rho{\mathcal{K}_{\vec{k}}}^\dagger,
\label{rho_noisy}
\end{equation}
where $\mathcal{K}_{\vec{k}}$ denotes the tensor product of certain combinations of the following two Kraus operators corresponding to the amplitude damping noise model
\begin{equation}
 K_0 =
 \begin{pmatrix}
  1 & 0\\
  0 & \sqrt{1-p} \\
 \end{pmatrix}
;\quad K_1 =
 \begin{pmatrix}
  0 & \sqrt{p} \\
  0 & 0\\
 \end{pmatrix},
\label{ampdamp}
\end{equation}
where $p$ stands for the probability that an excitation is lost. With these, $\mathcal{K}_{\vec{k}}$ is defined by
$\mathcal{K}_{\vec{k}} = \otimes_{l=1,\ldots,n}K_{k_l}$, where $n$ is the number of qubits and $k_l$ can take 0 or 1.

Using the W state, $\rho_W=|W_n\rangle\langle W_n|$ as the initial
state $\rho$ in~(\ref{rho_noisy}), the following relations turn
out to hold true:
\begin{align}
\mathcal{K}_{0\ldots 0}\rho_W\mathcal{K}_{0\ldots 0}^\dagger &= (1-p)\rho_W\nonumber\\
\mathcal{K}_{0\ldots 01}\rho_W\mathcal{K}_{0\ldots 01}^\dagger &= (p/n)|0^{\otimes n}\rangle\langle 0^{\otimes n}|,
\label{Kperm}
\end{align}
Further, if we permute the index $\vec{k}={0\ldots 01}$ in all the $n$ different ways, we get
the same state as that on the right hand side of the second line of~(\ref{Kperm}), i.e., the $n$-partite vacuum state multiplied by $p/n$. On the
other hand, if the number of $1$'s in $\vec{k}$ are at least two,
we have $\mathcal{K}_{\vec{k}}\rho_W\mathcal{K}_{\vec{k}}^\dagger
= 0$. Substituting into (\ref{rho_noisy}), we arrive at
\begin{equation}
\rho_{noisy}=(1-p)|W_n\rangle\langle W_n| + p|0^{\otimes
n}\rangle\langle 0^{\otimes n}|,
\end{equation}
which is the same state as (\ref{rhonp}).


\end{document}